\documentclass{article}

\usepackage[greek,english]{babel}
\usepackage{dirtytalk}
\usepackage{multicol}

\usepackage[letterpaper,top=2cm,bottom=2cm,left=3cm,right=3cm,marginparwidth=1.75cm]{geometry}

\usepackage[backend=biber,style=numeric,citestyle=numeric-comp]{biblatex}
\usepackage{hyperref}
\hypersetup{colorlinks,allcolors=blue}

\usepackage{comment}
\usepackage{amsmath}
\usepackage{amssymb}
\usepackage{graphicx}

\title{Refuting the Metaphysics of Wolfram and Tegmark}
\author{Joseph Natal}

\usepackage{setspace}
\begin{document}
\maketitle

\begin{abstract}
Wolfram's hypergraph dynamics should replace outmoded models in physics. This should even more so be the case if experimental evidence for the theory is found (which I believe is probable). However, due to the breadth and depth of the theory, it may be difficult to produce experimental evidence which \textit{falsifies} it. Some of Wolfram's personal work relating to his physics project is philosophical, and so mechanics of particular phenomena in the natural world can become a triviality or an aside. In other words, the general theory ``casts a wide net'', and it is the philosophical topics I will challenge. I find that Wolfram must adopt a radical epistemology through his so-called Observer Theory because there is no clear notion of Truth. Tegmark believes in an objective Truth, but I find its relation to the observer untenable, and the proof of his Mathematical Universe Hypothesis (MUH) is gematria. I argue both Wolfram and Tegmark conflate \foreignlanguage{greek}{(έφήρμοττον)} the inherent potential \foreignlanguage{greek}{(δύναμις)} of mathematical truths with their instantiation or actuality\foreignlanguage{greek}{
(ἐνέργεια)} in reality, making a similar error to that of the ``so-called'' Pythagoreans rebuked by Aristotle. Nonetheless, I believe that combinatorial structures of the kind used in the physics project (abstract rewriting, directed acyclic graphs) will be the future of physics as we know it.
\end{abstract}

Enlightenment English philosophy characterizes God as progressively more impersonal, with ``Nature'' and ``Law'' acting as substitutions for the imminent Creator \cite{darwin1859origin, DeclarationofIndependence:atranscription_2024}:
{
\newline
\newline
\centering
\textit{But with regard to the material world, we can at least go so far as this---we can perceive that events are brought about not by insulated interpositions of Divine power, exerted in each particular case, but by the establishment of general laws.}
\newline---Anglican William Whewell quoted on the first page of Darwin's Origin of Species 
\newline}

{
\centering
\textit{ \ldots and equal station to which the Laws of Nature and of Nature's God entitle them \ldots}
\newline---Thomas Jefferson in the Declaration of Independence, a deist and nominally a member of the Church of England 
\newline
\newline
}
\noindent
and Stephen Wolfram himself writes \cite{Howwegotherewolfram}:
{
\newline
\newline
\centering
\textit{As the centuries went by, the idea of ``natural laws'' sharpened, sometimes with an almost computational feel. ``God can only run the world by natural laws'', or ``The universe is the thoughts of God actualized'' }
\newline
\newline
}
So---perhaps influenced by the secular milieu of mid-to-late 20th century science---he develops this idea further: God did not give rules, God \textit{is} rules! And by doing so, it would seem that he completes the removal of anthropocentrism in creation and further divides what Heisenberg calls the Cartesian partition (``God-World-I'') \cite{Heisenberg}. But this assessment is not correct, because the dependence of mathematical truths on the experience of ``observers like us"(\textit{sic}) is a major conclusion of the theory \cite{Somehistoricalbackgroundwolfram}:
{
\newline
\newline
\centering
\textit{The intellectual foundations and justification are different now. But in a sense our view of mathematics has come full circle. And we can now see that mathematics is in fact deeply connected to the physical world and our particular perception of it. And we as humans can do what we call mathematics for basically the same reason that we as humans manage to parse the physical world to the point where we can do science about it. 
}
\newline
\newline
}
\indent How far does ``our particular perception'' allow us to speak on fundamental theories of physics and metaphysics? According to Wolfram, the underlying substrate of reality is the ``entangled limit of all possible computations'' called ``the ruliad''. This conjecture is either as abstract as the unknowable God and unfalsifiable, or so vacuous one might as well say ``the universe does things'', or some superposition of the two. Max Tegmark, on the other hand, asserts reality is all ``mathematical structure''---a form of radical platonism that aligns with his acceptance of the Many-Worlds interpretation of quantum mechanics \cite{Tegmark_1998}. Yes, a culmination of phenomena including quantum mechanics, the modern computer, and lack of alternative dogma have enabled what I might describe as gnostic atomism. The following paragraph will explicate a brief history of similar ideas, although a more in-depth history is found in Heisenberg's book on the philosophy of physics \cite{Heisenberg}.

\indent Democritus is the most well-known earliest proponent of the theory of the atom around 400 BC (who, like Wolfram and Tegmark, was controversial) and his theory is roughly aligned with modern atomic theories of today \cite{AncientAtomism}. In some ways modern theories supersede those of the past, but in other ways they are no different. For example, Aristotelian and Asharite philosophers described reality in terms of atoms and ``accidents'' \cite{ArabicandIslamicNaturalPhilosophy}, Democritus described atoms and ``hooks'' \cite{Democritus} which are both similar to the idea of vertices and edges on Wolfram's hypergraphs.  It is unsurprising that Wolfram's use of natural philosophy and transcendent ruliad bear similarity to Islamic ideas. Abu al-Hudhayl---regarded as the founder of kalām atomism---posited that atoms were indiscernible from one another in size or shape, similar to emes, Wolfram's fundamental atom of computation. Emes are given unique identifiers, which is comparable to Leibniz's Monads where ``each Monad must be different from every other'' and he describes them as ``incorporeal automata'' \cite{Leibniz_1714}. While atomism of the past relied on crude arguments using primitive technology such as grain mills \cite{Dhanani_2015} or steam engines, Wolfram and Tegmark use the digital computer to animate their ideas. The advantage of the modern computer for Wolfram's theory is that it allows for large combinatorial statistics experiments to be run, which form the basis for reproducing all natural laws of 20th century physics that physicists care about. More recent developments for discrete models of our universe come from Rafael Sorkin's causal set theory \cite{causalset} and Ed Fredkin's SALT cellular automaton model. Before them, John Von Neumann explored self-replicating cellular automata with the goal of designing a machine whose complexity could grow automatically akin to biological organisms under natural selection \cite{neumannautomata}.

In this work I will undertake a similar project to philosopher of science Nancy Cartwright who aligns with me in the ``Aristotelian belief in the richness and variety of the concrete and particular'' \cite{cartwrightlies} and will reframe some of the challenges put forward by Aristotle in \textit{Metaphysics} \cite{metaphysicsaristotle}. It was Aristotle who criticized both Plato and fourth-century Pythagoreans for constructing natural bodies possessing weight from indivisible mathematical abstractions \cite{Pythagoreanism}. An error of the ``so-called'' Pythagoreans is described in \cite{platopythag}:
{
\centering
\newline
\newline
\textit{The ``so-called'' Pythagoreans adapt the immediate facts to fit the explanation in an ad hoc manner. The example given involves the bodies of the heavens: nine bodies can be perceived by the senses, but, since the Pythagoreans assume the number 10 as the perfect number, and since all things are number, there must be ten heavenly bodies. This example also reveals Aristotle's second substantial criticism of the philosophical system of the ``so-called'' Pythagoreans: they hastily and carelessly compare things in order to secure relationships between their first principles and observed phenomena.}
\newline
}
\newline
\newpage
\indent We begin by discussing the current assertions given by Wolfram himself. In the model we assume observers like us have \cite{ObserverTheory—StephenWolframWritings_2023}
\begin{enumerate}
\centering
	\item Persistence in time 
	\item Persistence in space 
	\item ``Independence'' from the universe
	\item Stability of discrete symbolic concepts
	\item Independence from the hyperruliad
\end{enumerate}

Tegmark shares the first four beliefs. Some assertions of Wolfram regarding these assumptions are as follows:

{
\textit{
\begin{itemize}
\centering
\item But why should that summary (the perception of our universe) have any coherence? Basically it's because we impose coherence through our definition of how observers like us work.\\
\item But actually there’s more than just our assumption of persistence in time. For example, we also have an assumption of persistence in space: we assume that—at least on reasonably short timescales—we’re consistently ``observing the universe from the same place'', and not, say, ``continually darting around''.\\ 
\item Because in fact observers that are even vaguely like us are in effect deeply constrained in what rules they can attribute to the universe.\\
\item But we make the implicit assumption that, no, the universe at least as far as we perceive it - is a more organized and consistent place. And indeed it's that assumption that makes it feasible for us to operate as observers like us at all, and to even imagine that we can usefully reduce the complexity of the world to something that ``fits in our finite minds''. \\
\item We assume that we can take all the complexity of the world and represent at least the parts of it that we care about in terms of discrete symbolic concepts, of the kind that appear in human (or computational) language.\\
\item These assertions seem very general, and in some ways almost self-evident-at least as they apply to us. \\
\end{itemize}
}
}

In using the plural ``us'' and ``our'' consistently, we can assume Wolfram is assuming a common agreement amongst readers, or is employing nosism. Either way, the presumption that many humans have a common sampling of the ruliad is left unexplained, along with why we are and remain localized in space and time. It is unclear how observers are delineated in the ruliad and how this is compatible with the purported independence from the universe---the idea of ``observers like us'' is simply a brute fact for Wolfram. These proposed restrictions on observers bring to mind the fine-tuning argument for intelligent design: as observers, we happen to be \textit{just the right kind} to be able to name the underlying structure of reality itself. To quote Wolfram, ``There has to be just enough equivalencing'' of information by human observers ``and not too much.'' He admits circularity in his reasoning himself:
{
\newline
\newline
\centering
\textit{
Well, the point is that an observer can equivalence those successive patterns of emes, so that what they observe is persistent. And, yes, this is at least on the face of it circular. And ultimately to identify what parts of the ruliad might be ``persistent enough to be observers'', we'll have to ground this circularity in some kind of further assumption. 
}
\newline
\newline
}

\begin{figure}[!h]
    \centering
    \includegraphics[width=0.5\linewidth]{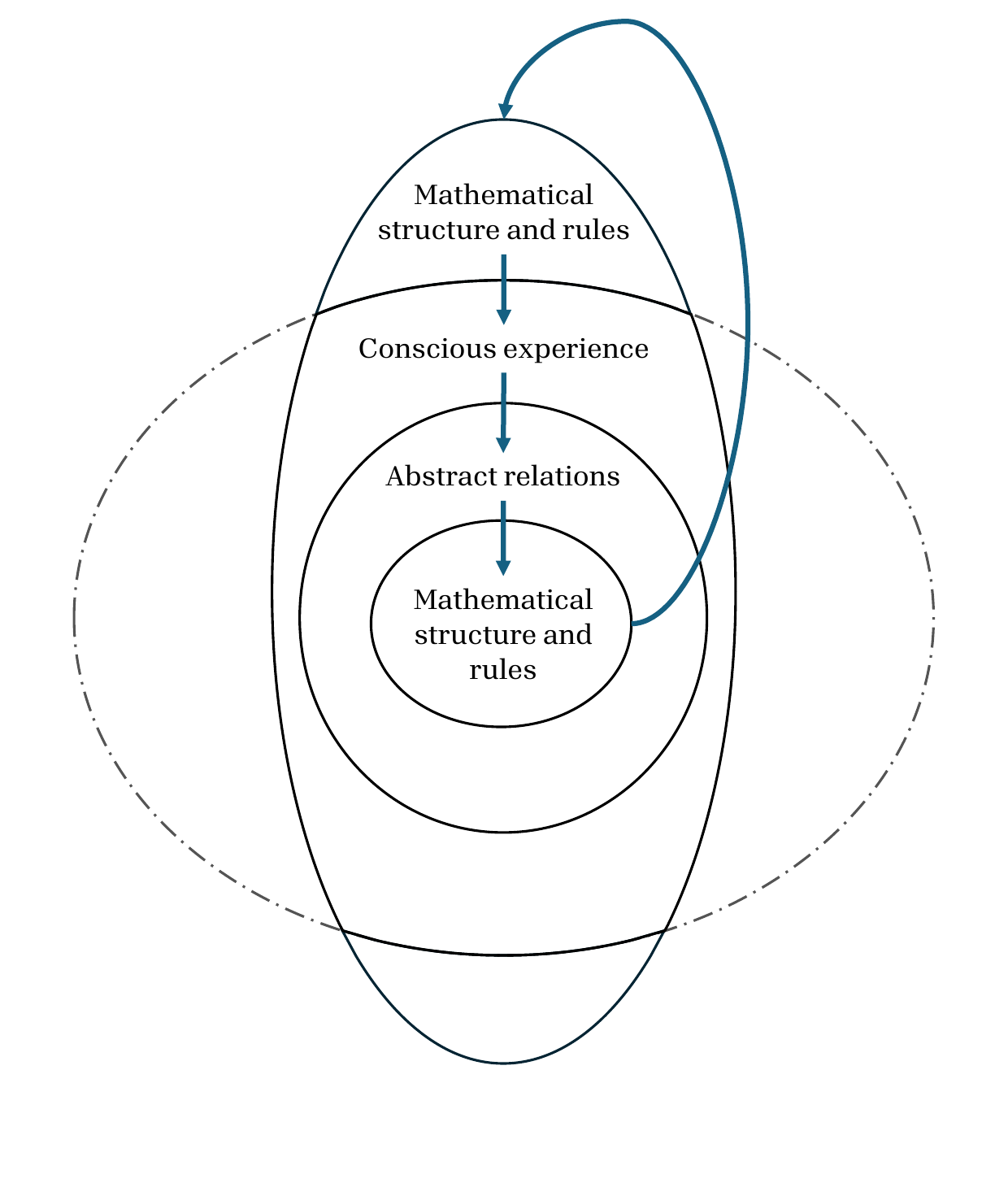}
    \caption{An interpretation of the metaphysics given by Wolfram's ruliad and Tegmark's MUH that exhibits circularity.}
    \label{fig:enter-label}
\end{figure}

So Wolfram's current universe is dyadic---it consists of the ruliad and observers. If conscious observers were emes, these emes would still follow privileged rules which uphold the aforementioned assumptions regarding the observer. For example, it could be that observers are time-and-space-localized compression emes which sample the ruliad. But, in seeming contradiction, Wolfram (and Tegmark) are notably unsatisfied with a privileged rule which runs the universe:
{
\newline
\newline
\textit{
At the outset, we might have imagined that the end point of our project would be the identification of some particular rule of which we could say ``This is the rule for the universe''. But of course then we'd be faced with the question: ``Why that rule, and not another?'' And perhaps we would imagine just having to say ``That's something that you have to go beyond science to answer''.}
\newline
\newline
}
Wolfram and Tegmark avoid this challenge like the ``so-called'' Pythagoreans in the aforementioned example by claiming that instead of one rule, \textit{all} rules are run. Wolfram forces the mathematical physics jargon ``in the entangled limit''. I suspect by invoking terms used often in quantum mechanics and calculus, the reader should naturally intuit what he means.
\newline \indent Let's turn to more technical definitions of the observer as given by Gorard \cite{gorard1}:
{
\newline
\newline
\textit{$\hdots$By defining the observer mathematically as a discrete hypersurface foliation of the multiway evolution graph$\hdots$}
}
{
\newline
\newline
\textbf{Definition 24} \textit{An “observer” is any persistent structure within a multiway system that perceives a single,
definitive evolution history.}
\newline
}

{

\noindent \textbf{Definition 37} \textit{An “observer” in a multiway system is any ordered sequence of non-intersecting branchlike
hypersurfaces $\Sigma t$ that covers the entire multiway graph, with the ordering defined by some universal time
function...}
\newline
\newline
}
Critical to the model is the abstract rewriting system \cite{gorard2}:
{
\newline
\newline
\textbf{Definition 5} \textit{ An ``abstract rewriting system'' is a set, denoted A (where each element of A is known as an ``object''), equipped with some binary relation, denoted $\rightarrow$, known as the ``rewrite relation''}. 
\newline
\newline
}
\indent How do we corroborate Wolfram's assertions with Gorard's definitions and our conscious experience? From the definitions the model of reality is \textit{relations} between vertices on hypergraphs. A binary \textit{relation} is defined as a set of ordered pairs of elements from two sets. So from the eternally existing atoms of space such relations spring forth. Now these relations must be effectively acting at distances like gravity on celestial bodies. Presumably due to their fundamental nature these interactions are inscrutable unless seen by secondary effects. However, once we perform experiments we are thrust into a world of instances, we can weave any narrative we want regarding what we perceive on the detectors, 
and despite being flexible enough for fitting, the model is vacuous because \textit{which} rules applied \textit{when} presumably must evolve according to some hyperrule---the ruliad is a secondary creation. Regarding the so-called ``hyperruliad'' existing, Wolfram writes the following \cite{Wolfram_ruliad}:
{
\newline
\newline
\textit{
Yes, we could imagine some other entity that's embedded within the hyperruliad, and perceives what it considers to be the universe to operate hypercomputationally. But in a statement that's in a sense more ``about us'' than ``about the universe'', we assert that that can't be us, and that we in a sense live purely within the ruliad---which means that for us the Principle of Computational Equivalence holds, and we perceive only computation, not hypercomputation.
}
\newline
\newline
}
The is reminiscent of the question of the creator of God---``If God exists then who created God?''. Wolfram's rather unsatisfactory answer in the context of the hyperruliad existing is basically ``we assert we are not in the hyperruliad because it's not what we perceive'' (presumably speaking on behalf of humankind).

Perhaps this problem can be solved if Aristotle's challenge to the Pythagoreans is answered \cite{metaphysicsaristotle}:
{
\newline
\newline
\textit{Those who say that existing things come from elements and that
the first of existing things are the numbers, should have first
distinguished the senses in which one thing comes from
another, and then said in which sense number comes from its
first principles.}
\newline
\newline
}
Even granting abstractions without agency like numbers the ability to create instances, I will further show that observer-independent truth is impossible and so claims about external reality are void unless there is an explicit reconciliation. Consider the following objections a proponent of the theory might state:


\noindent \textit{O: hypergraphs don't actually compose the ruliad}\\
\indent If not then there is a great disconnect between the models of the theory and the theory itself.\\
\noindent\textit{O: The ruliad itself is an abstraction---an abstraction of all transformations on hypergraphs}\\
\indent And so its instantiation is an article of faith. Then suppose Jonathan, a skeptic, disagrees on the grounds that the assumptions are not self-evident. Perhaps he is a determinist and rejects the claim we are ``independent'' from the universe. Using the model and language of Wolfram:
\newpage
\begin{multicols}{2}
\centering
\small
Stephen is an observer like us.\\
Stephen performs a lossy compression on a local space of rules adjacent to ours.\\
Stephen claims that $x$ is self-evident\\
\vfill
\columnbreak
Jonathan is an observer like us.\\
Jonathan performs a lossy compression on a local space of rules adjacent to ours.\\ 
Jonathan claims that $x$ is $\lnot\;$self-evident\\
\end{multicols}

\noindent\textit{O: Stephen and Jonathan can clear up their disagreement by proving x with a computational language}\\
\indent For problems in non-computational fields, Wolfram does not claim to have resolutions (but will soon!).\\
\textit{O: Grant that all the problems in our universe can be resolved by a computational language}\\
\indent In the model, \textbf{truth is an emergent artifact of our nature as observers} \cite{Wolfram_truth}:
{
\newline
\newline
\noindent \textit{
So what happens if one tries to add a statement that “isn’t true”? The basic answer is that it produces an “explosion” \ldots From the point of view of underlying rules—or the ruliad—there’s really nothing wrong with this. But the issue is that it’s incompatible with an “observer like us”—or with any realistic idealization of a mathematician.}
\newline
\newline
}
Because false statements and contradictory axioms exist in the ruliad, we must fall back and appeal to our nature, leaving the above disagreement up to some ill-defined consensus. Wolfram, by making the assumption of the ``stability of discrete symbolic concepts of the kind of a computational language'', contradicts himself unless he can give an account for the faculty of abstract representation and its stability which doesn't rely on the argument of usefulness from evolution (this would undermine its objectivity). But this is impossible because unlike in Tegmark's universe, mathematical Truth is observer-dependent, and so perhaps disagreement---and moreover logic---is a kind of ambient babble.\\
\indent In brief, we take the statement ``the ruliad exists'' as having a Truth value, but truth is according to our nature, and if it's according to our nature, it's not according to external reality and so does not have a Truth value---a contradiction.\\
\noindent \textit{O: Any human would find the ruliad reasonable given the common-sense assumptions}\\
\indent I don't believe an appeal to common belief is proof, and this argument relies on the same anthropocentrism that Wolfram and Tegmark desperately try to avoid. External intelligence is also proposed in the model, so in the above example Jonathan can be replaced with an alien, or a rock. Is the ruliad no longer an objective concept if an alien mind could not have a conception of this ruliad? To go further, what if \textbf{our nature as ``observers like us'' give rise to the abstraction of the ruliad}? 
\\ \noindent \textit{O: Our nature as observers does give rise to the abstraction of the ruliad}\\
\indent Then the ruliad is \textit{about us}. Then it can be true, false, or anything in between, if all is merely computation. At some point one would expect an extrapolation to reality independent of observers.\\ 
\noindent \textit{O: These are definitions of the observer which reproduce 20th century physics}

Reproducing 20th century physics is reproducing models, not revealing the underlying structure of reality. And despite the purported potency of explaining the motion of bodies, Cartwright and Van Fraassen ask ``What is it about explanation that guarantees truth?". That being said, with the flexible combinatorial nature of the model it will likely be the case that many amazing phenomena will be able to be described by it.

\noindent \textit{O: The ruliad exists by formal necessity}\\
\indent But isn't ``formality'', a word also used in mathematics as well as philosophy, laden with an observers presuppositions? Gorard describes Definition 5 as coming from a formalism within mathematical logic, so it appears mathematical logic subsumes graph rewriting systems, and binary relations have an \textit{a priori} existence undergirding them. Is there a causal chain connecting these concepts and how was this knowledge imprinted onto us particular observers? In Plato's case, knowledge of his theory was recollection from the eternal soul. David Hilbert addressed mathematical theory as relations in \cite{Blanchette_2007}:
{
\newline
\newline
\textit{
It is surely obvious that every theory is only a scaffolding or schema of concepts together with their necessary relations to one another, and that the basic elements can be thought of in any way one likes. If in speaking of my points I think of some system of things, e.g., the system: love, law, chimney-sweep $\hdots$ and then assume all my axioms as relations between these things, then my propositions, e.g., Pythagoras' theorem, are also valid for these things. In other words: any theory can always be applied to infinitely many systems of basic elements. 
}
\newline
\newline
}
Wolfram has then applied this to his theory of relations between emes, the intangible atoms of computation. \textit{Relation} comes from the Latin word referre which means ``to carry back, notify, report'', which reminds us of colloquial usage when we ``relate to'' someone a message (it is also the root word for referee, referrant). I bring this up to frame the following question: how can one prove ontological priority of causal relations between emes over other relations (e.g. conferring a message between two humans, an emotional relation between two humans, a qualitative relation between colors)? Somehow, emes have privileged existence, otherwise they can be confidently relegated to the rest of the ``coarse-grained'' observations which comes from observers like us. I reject the argument that emes necessarily exist for there to be any kind of regularity or predictability in the universe due to insufficient evidence. 

\begin{figure}[!h]
    \centering
    \includegraphics[width=\linewidth]{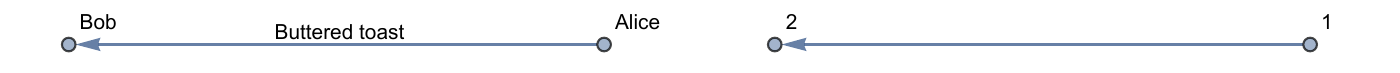}
    \caption{A causal graph laden with composite human-like ideas and a more abstract causal graph of unique identifiers. Which is which? }
    \label{fig:enter-label}
\end{figure}

\newpage
\noindent\textit{O: The ruliad exists in the same way 1+1=2 exists}\\
\indent The ruliad then also exists in the same way $1+1=3$ exists because truth is an emergent artifact of observers, and no formal axiom system is given precedence in the ruliad. Wolfram must explain the privilege of $1+1=2$ in some observer-independent way. Supposing he can do this, I believe I agree with the claim. However, I do not agree that $1+1=2$ implies that reality \textit{is} the ruliad (but perhaps this can be proved with a presupposed ``stable'' formal language and an automated theorem prover!). The main error I see is the conflation of the potential \foreignlanguage{greek}{(δύναμις)} of $1+1=2$ with the actuality \foreignlanguage{greek}{(ἐνέργεια)} of placing two rocks together in an instantiation. $1+1=2$ also does not \textit{cause} two rocks to be placed together.
\begin{align*}
\text{truths \;\;} &\;\;\vert \text{\;\;\;\; instances}\\
1 + 1 = 2 &\nRightarrow \text{two rocks exist}\\
\therefore \text{the ruliad} &\nRightarrow \text{the ruliad is reality}
\end{align*}
No, Wolfram would concede it is $\text{ruliad} + \text{observer theory} + \text{observers} = \text{reality}$. Two rocks together exist independently of whether or not $1+1=2$ exists and $1+1=2$ exists without placing two rocks together. So I ask Wolfram which is more true in his mind: ``1+1=2'' or ``I am conscious''? The latter of the two statements represents a fundamental empirical or instantial truth (of the observer), while the former a fundamental transcendental truth (of the ruliad). Is there a contingency or priority between those two truths?

\begin{figure}[!h]
    \centering
    \includegraphics[width=0.5\linewidth]{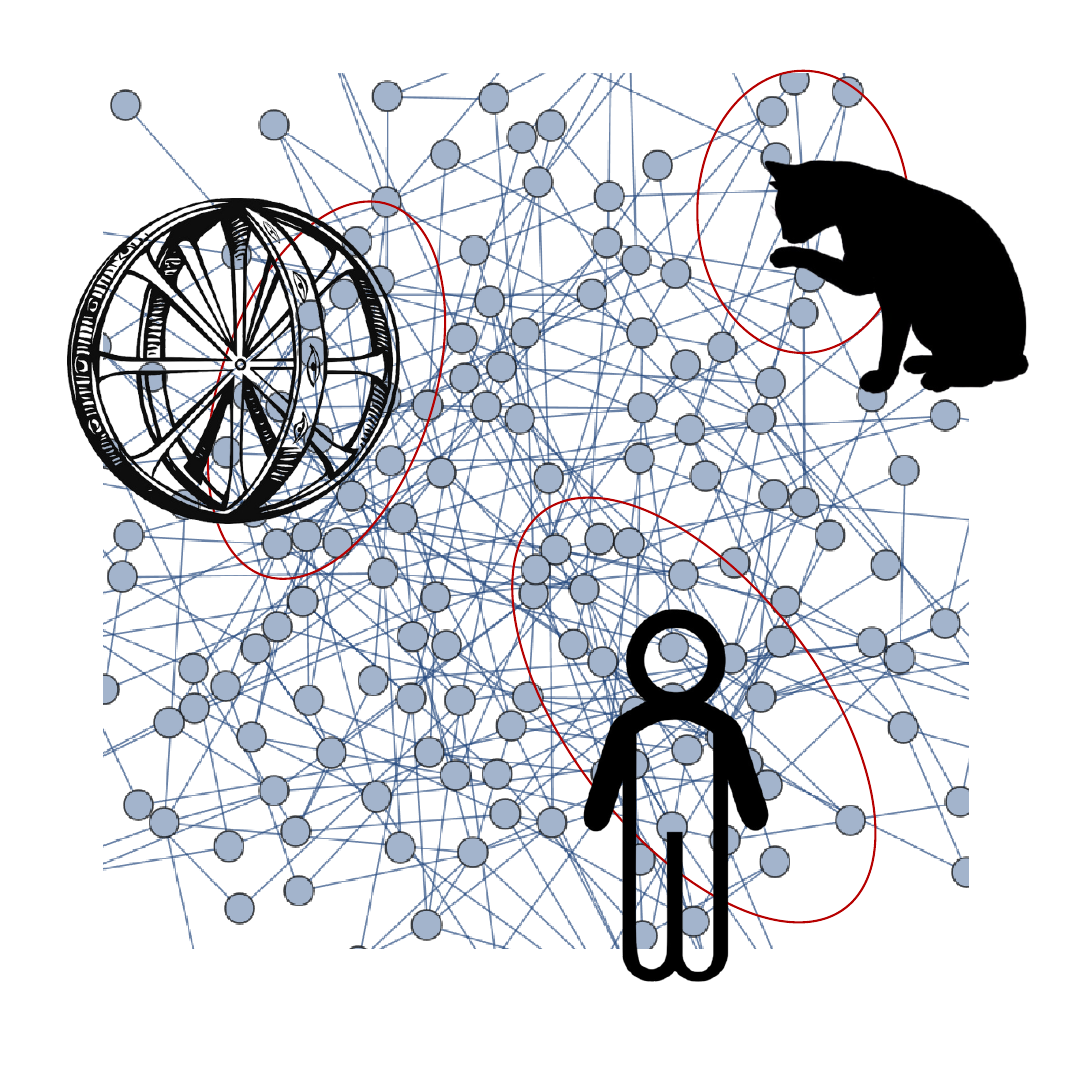}
    \caption{Candidate rulial observers sampling the ruliad. As an observer one might reject one of these candidates, but hopefully the \textit{true} observers  will be known once the theory is complete.}
    \label{fig:enter-label}
\end{figure}

An evolutionary biologist could claim that some humans may have been selected for based their ability to control external bodies in time and space, and to communicate. Then one can see how giving elevated---almost deified---status to the fundamental object (atoms) or to fundamental language (formal language) could be a byproduct of our evolution. Perhaps given this reductionist perspective a ``great adjacency matrix in the sky'' is a reasonable conclusion.

I will now briefly address some of Tegmark's claims, and in doing so expound on some arguments already addressed above.  Here are some key definitions given by Tegmark \cite{Tegmark_book}:
{
\newline
\newline
\textbf{Mathematical Universe Hypothesis (MUH)}: \textit{Our external
physical reality is a mathematical structure.}
\newline
\newline
}
{
\centering
\textbf{Mathematical structure}: \textit{Set of abstract entities with relations
between them.}
\newline
\newline
}
{
\textbf{Equivalence}: \textit{Two descriptions are equivalent if there’s a correspondence between them that preserves all relations.}
\newline
\newline
}

Tegmark's definition of equivalence only equates \textit{descriptions}, and I posit that it would sound more absurd if instead the word \textit{objects} was used. His language is vague by necessity---if someone told me the universe was \textit{abstract entities with relations between them} I would assume this person was speaking about the spiritual realm. His intangible entities, like souls, also happen to be eternal. Furthermore his definition of equivalence is exactly that---it is not equality.\\

{
\textit{... If you can
thus pair up every entity in our external physical reality with a corresponding one in a mathematical structure (“This electric-field strength here in physical space corresponds to this number in the mathematical structure,” for example), then our external physical reality meets the definition of being a mathematical structure---indeed, that same mathematical structure}
\newline
\newline
}
\begin{figure}[h]
    \centering
    \includegraphics[width=0.5\linewidth]{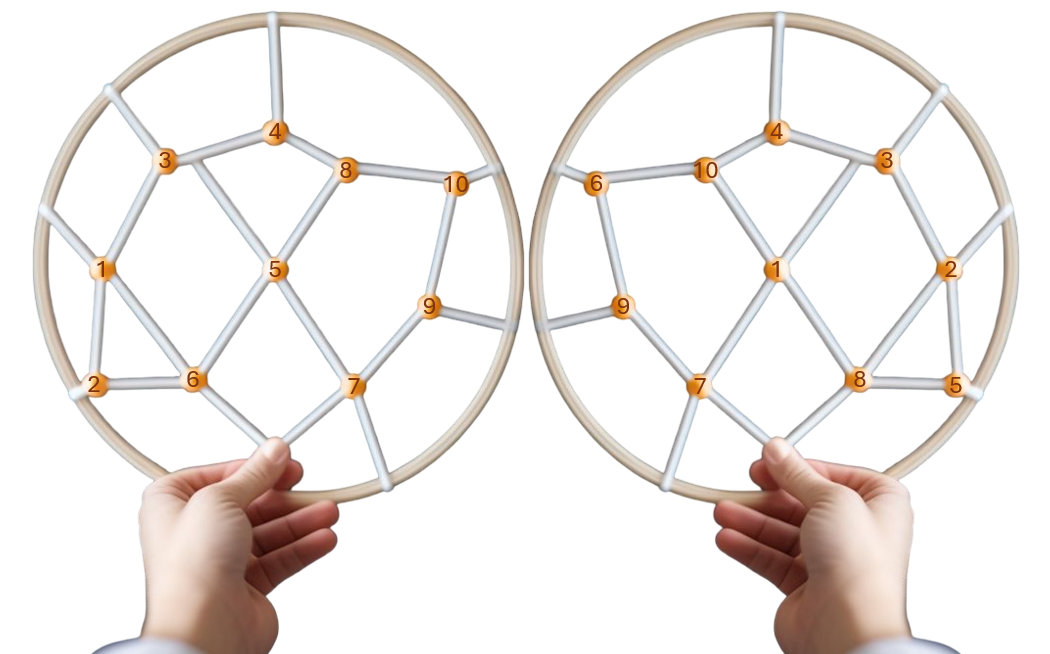}
    \includegraphics[width=0.501\linewidth]{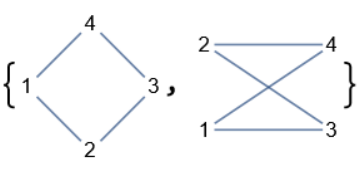}
    \caption{
    There is a \textit{twoness} inherent in both figures: in the top figure, the physical matter held in each hand is not identically equal and in the bottom figure---supposing a multiset---the cardinality is two. There is also a \textit{oneness} in that relations between abstract components are preserved---the graphs are isomorphic. I believe both properties are real and eternal, but asserting \textit{oneness} presupposed \textit{twoness} and required subsequent computation. 
    }
    \label{fig:isographs}
\end{figure}
\indent Tegmark's proof reduces to checking a lookup table of abstract models to physical measurements, which he expresses as a graph isomorphism problem (Fig. \ref{fig:isographs}). This seems rather contrived and not my definition of identically equal: by using the phrase ``these two things are identical'' one already presupposes distinctness with the word \textit{two}. Then one has to prove how his or her perception and process of enumeration failed to identify their \textit{oneness}, rather than claiming they are identically equal because their respective components can be paired.

This aside, I assert that his method of proof places an impossible burden of computation on the observer and doesn't escape the measurement problem. Tegmark's theory radically diverges from Wolfram's here in that Tegmark would need an observer that is not \textit{computationally bounded}, a term used by Wolfram to describe the fact that despite knowing the rules or equations, we as humans can't immediately predict certain futures. In spite of Tegmark's view of an ultimately static universe, he still claims that the act of \textit{pairing up} is required to prove his hypothesis. Then the question is: who is doing the pairing up, when, where, and how often? Graph isomorphism may be NP-complete and the number of vertices in this problem seems infinite. Tegmark acknowledges an internal reality subject to hallucinations, and how this is circumvented is unclear. Consider Dirac's speculation \cite{dirac}:
{
\newline
\newline
\textbf{Dirac's speculation:} The constants of nature change over time $\approx$ the age of the universe. 
\newline
\newline
}
\noindent This can still be consistent with a universe made of mathematics, but it confounds the claim that the math we have access to \textit{now}, such as the Schrödinger equation, represents the immutable and eternal. There are always higher order corrections that can be woven into the narrative. To take a step back, \textit{pairing up} requires the notion of binary relations and relation in mathematics is defined in the abstract. This, in and of itself, is not objectionable, but an account for the apparent and emergent faculty of humans to abstract is needed as \textit{abstract relations} undergird the entire theory. So what does Tegmark mean by \textit{relation} in his definition of mathematical structure? I would describe Max as an \textit{abstract entity} with whom I have a \textit{relation}. I would describe souls as \textit{abstract entities} with \textit{relations} between other souls, but although semantically correct, this is likely not within the scope of Tegmark's definition.

\noindent \textit{O: The relations that describe Max are, in a first approximation, ``human'', which is not the case for math abstractions.}\\
\indent Maybe Tegmark believes his human-independent description of the world, but as this theory comes from a human, I question the idea. In fact, abstraction is one of the \textit{most} human things one can do. But I suppose a more apt description for Tegmark is a self-described ``mathematical pattern in spacetime''. From Tegmark: 
{
\newline
\newline
\textit{it must also be well defined according to nonhuman entities aliens or supercomputers, say—that lack any understanding of human concepts. Put differently, such a description must be expressible in a form that's devoid of any human baggage like ``particle'', ``observation'' or other English words.}
\newline
\newline
}
\indent or from Wolfram:
{
\newline
\newline
\textit{So we need to get rid of the “human-readable names” and just replace them with “lumps” $\hdots$ }
\newline
\newline
}
What baggage is in Tegmark's definition of ``baggage'' and how does one conceive of ``lumps'' without instances? Because the computer is the tool du jour, I'll make an analogy and say that this is like a class without an instance, or compile-time without run-time. Tegmark illustrates an example of non-humanness by comparing the description of a chess game using English language to one of algebraic chess notation, but it seems like an arbitrary choice to ascribe a more fundamental existence to the algebraic chess notation over, say, the narrative of the chess game. 
\begin{figure}[!h]
    \centering
    \includegraphics[width=0.9\linewidth]{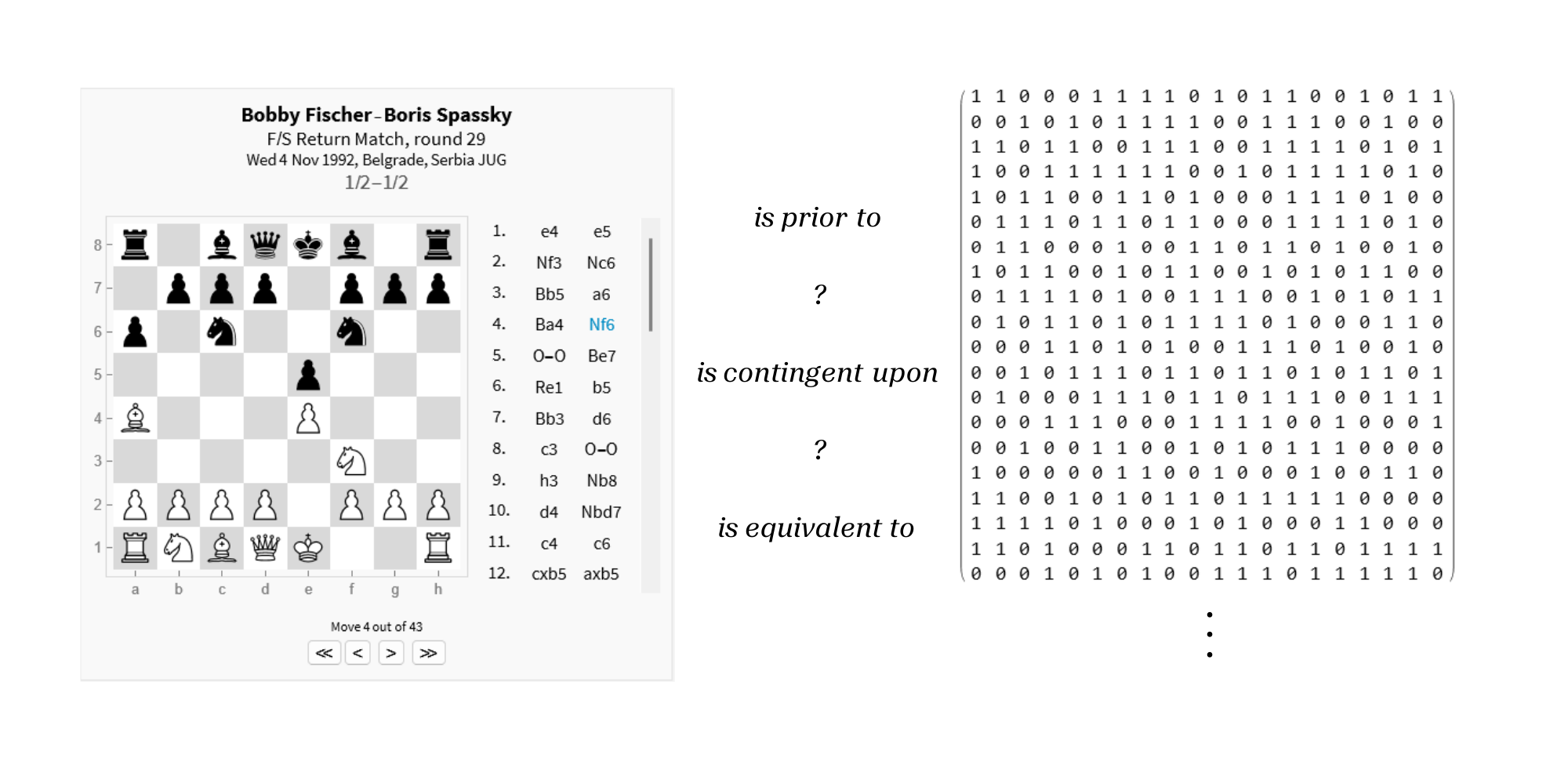}
    \caption{A chess game and its compressed Forsyth–Edwards notation, converted to a byte string and then a binary number \cite{Chojna_Warendorff_2023}. Wolfram and Tegmark think the binary notation is something more like the fabric of reality, whereas a chess player might want to look at the chess board and the players.}
    \label{fig:chess}
\end{figure}

\noindent \textit{O: It is obvious that mathematical relations are significantly less human than, say, love between people}\\
\indent I personally cannot touch, feel, or perceive pure human love but I believe it exists in a similar platonic way that $1+1=2$ exists. It must be that Maxwell's demon is picking and choosing what is allowed in our platonic realm.\\
\noindent \textit{O: Human love doesn't predict motion of bodies in space and time}\\ \indent Prediction is a poor stand-in for truth.\\
\noindent \textit{O: Aliens and humans would have the same math, unlike platonic love}\\
\indent Wolram may disagree, asserting that math is intimately linked with our physical experience as observers. Furthermore, Tegmark's supposed universal math has not been shown to be rigorous:
\newline
\newline
{
\noindent \textit{All of this begs the question: is it actually possible to find such a
description of the external reality that involves no baggage? If so, such
a description of objects in this external reality and the relations between
them would have to be completely abstract, forcing any words or symbols to be mere labels with no preconceived meanings whatsoever. Instead, the only properties of these entities would be those embodied by the relations between them.}
\newline
\newline
}
where the last sentence recapitulates Hilbert's definition of theory as relations. So how do we come to terms with individuated and connected nothings giving rise to our experience?

\noindent \textit{O: Relations are emergent phenomena}\\
\indent Indeed, Tegmark asserts a kind of emergence:
{
\newline
\newline
\textit{The external physical reality is therefore more than the sum of
its parts, in the sense that it can have many interesting properties while
its parts have no intrinsic properties at all.}
\newline
\newline
}
So after all the chapters describing the amazing reductionism in phenomena of physics (e.g. claiming that colors are actually electromagnetic spectra of Maxwell's equations, everything is built of atoms), Tegmark asserts that properties can \textit{ex-nihilo} emerge, but only from these particular ``propertyless'' entities. Would he grant such emergence for all other physics phenomena in opposition to their reductionist explanation? Along with his concept of \textit{perceptronium}, perhaps he should introduce \textit{emergetronium} as another distinct object, or else define emergence as a property, albeit latent, of mathematical structures.\\
\noindent \textit{O: These emergent properties are all artifacts of our consciousness}\\
\indent Tegmark believes consciousness is a ``feeling'', but I would argue ``feeling'' takes place in consciousness and consciousness is needed to describe feeling in the first place. 
He continues:
{
\newline
\newline
\noindent \textit{So if the entities of this structure have no intrinsic properties, does
the structure itself have any interesting properties (besides having eight
elements)? In fact, it does: symmetries!}
\newline
\newline
}
Now \textit{emergetronium} must be invoked as there are not only relations, but by introducing symmetry Tegmark describes transformations on relations, like those on Wolfram's hypergraphs. 

Let us assume a person was born blind, deaf, and overall without senses. Perhaps they have sight, but exist in what we would describe as an empty space. What is this person's understanding of symmetry, relations, feeling and enumeration? This person would only have a relation to ``I am'' which emerges from consciousness. Perhaps there's some biological hard-wiring which would give them conceptions of symmetry, and this may confound the claim that math is non-human. Tegmark explores the theory of the observer through this lens of biological evolution: 
{
\newline
\newline
\textbf{We perceive that which is useful}:\textit{
Why do we perceive the world as stable and ourselves as local and
unique? Here’s my guess: because it’s useful $\hdots$ Self-awareness would then
be a side effect of this advanced information processing.}
}\\ \\
but doesn't turn this reasoning about evolved usefulness on math itself. And in the reductionist evolutionary theory, self-awareness is a random mutation with \textit{no} evolutionary advantage, because an organism could be unself-aware but still perform all the same motor movements and speech for procreation. By the same token, self-awareness could have evolved to be shared or infinitely-spanning. Tegmark makes the same circular guesses as Wolfram regarding the locality and uniqueness of the observer.

In a response to the question ``Surely `stuff' can't be mathematical?'', Tegmark explains by making the analogy to a character in a video game who is not made of ``stuff'' and then writes \cite{Tegmark}:
{
\newline
\newline
\noindent \textit{Sure, the computer in this example is itself made of stuff, but the feeling that the objects in the game were made of ``stuff'' was completely illusory and independent of the substrate out of which the computer was built.}
\newline
\newline
}

\noindent I would like for him to think of an example that doesn't \textit{actually} have the ``substrate out of which the computer is built'', rather than to have to explain this caveat. Then I would be satisfied, knowing that abstractions without agency can act as a first cause and bring about instances. 

Wolfram clinically notes that ``the only thing special about the universe to us is us ourselves'' \cite{Wolfram_toe} and I believe Tegmark, who maintains a great interest in ethics, would agree. Wolfram modifies the Cartesian partition (``God-World-I'') by reintroducing ``I'' as ``observers'' and attempts to meld them with ``World''. Tegmark removes ``I'' and ``God'' and asserts that everything is in some sense ``World''. I understand why Heisenberg noted that modern science decided in favor of Plato and warned that the Cartesian partition is dangerous: in both theories of everything I find inconsistencies in the reconciliation of ``I'' the observer and the transcendent truths of mathematics. A cogent narrative has yet to be written which marries potential and instantiated truth. Aristotle thought that potentiality so understood is indefinable, claiming that the general idea can be grasped from a consideration of cases. Like Aristotle, I believe that actuality is to potential as \cite{aristotle}
{
\newline
\newline
\textit{“what is awake is in relation to what is asleep, and what is seeing is in relation to what has its eyes closed but has sight, and what has been shaped out of the matter is in relation to the matter”}

}

\printbibliography

@book{darwin1859origin,
  title={On the Origin of Species by Means of Natural Selection, Or, The Preservation of Favoured Races in the Struggle for Life},
  author={Darwin, Charles},
  lccn={06017473},
  url={https://books.google.de/books?id=jTZbAAAAQAAJ},
  year={1859},
  publisher={J. Murray}
}

@misc{Wolfram_truth, title={The notion of truth}, url={https://www.wolframscience.com/metamathematics/the-notion-of-truth/}, journal={The Notion of Truth}, author={Wolfram, Stephen}}

@misc{Wolfram_toe, title={Finally we may have a path to the fundamental theory of physics... and it’s beautiful-stephen wolfram writings}, url={https://writings.stephenwolfram.com/2020/04/finally-we-may-have-a-path-to-the-fundamental-theory-of-physics-and-its-beautiful/}, journal={Stephen Wolfram Writings RSS}, author={Wolfram, Stephen}, year={2020}, month={Apr}}

@misc{DeclarationofIndependence:atranscription_2024, author={Jefferson,Thomas}, title={The Declaration of Independence}, url={https://www.archives.gov/founding-docs/declaration-transcript}, journal={National Archives}, year={1776},  language={en} }

@misc{Howwegotherewolfram, title={How we got here: The backstory of the wolfram physics project-stephen wolfram writings}, url={https://writings.stephenwolfram.com/2020/04/how-we-got-here-the-backstory-of-the-wolfram-physics-project/}, journal={Stephen Wolfram Writings RSS}, author={Wolfram, Stephen}, year={2020}, month={Apr}}

@misc{Somehistoricalbackgroundwolfram, title={Some historical (and philosophical) background}, url={https://www.wolframscience.com/metamathematics/some-historical-and-philosophical-background/}, journal={Some Historical (and Philosophical) Background}, author={Wolfram, Stephen}}

@InCollection{AncientAtomism,
	author       =	{Berryman, Sylvia},
	title        =	{{Ancient Atomism}},
	booktitle    =	{The {Stanford} Encyclopedia of Philosophy},
	editor       =	{Edward N. Zalta and Uri Nodelman},
	howpublished =	{\url{https://plato.stanford.edu/archives/win2022/entries/atomism-ancient/}},
	year         =	{2022},
	edition      =	{{W}inter 2022},
	publisher    =	{Metaphysics Research Lab, Stanford University}
}

@misc{Wolfram_ruliad, title={The concept of the ruliad-stephen wolfram writings}, url={https://writings.stephenwolfram.com/2021/11/the-concept-of-the-ruliad/}, journal={Stephen Wolfram Writings RSS}, author={Wolfram, Stephen}, year={2021}, month={Nov}}

@book{Leibniz_1714, title={The Monadology}, url={https://homepages.uc.edu/~martinj/History_of_Logic/Leibniz/Leibniz%20-%20Monadology.pdf}, author={Leibniz, Gottfried Wilhelm}, editor={Latta, Robert}, year={1714} }

@InCollection{ArabicandIslamicNaturalPhilosophy,
	author       =	{McGinnis, Jon},
	title        =	{{Arabic and Islamic Natural Philosophy and Natural Science}},
	booktitle    =	{The {Stanford} Encyclopedia of Philosophy},
	editor       =	{Edward N. Zalta},
	howpublished =	{\url{https://plato.stanford.edu/archives/spr2022/entries/arabic-islamic-natural/}},
	year         =	{2022},
	edition      =	{{S}pring 2022},
	publisher    =	{Metaphysics Research Lab, Stanford University}
}

@InCollection{Democritus,
	author       =	{Berryman, Sylvia},
	title        =	{{Democritus}},
	booktitle    =	{The {Stanford} Encyclopedia of Philosophy},
	editor       =	{Edward N. Zalta and Uri Nodelman},
	howpublished =	{\url{https://plato.stanford.edu/archives/spr2023/entries/democritus/}},
	year         =	{2023},
	edition      =	{{S}pring 2023},
	publisher    =	{Metaphysics Research Lab, Stanford University}
}

@article{Dhanani_2015, title={The Impact of Ibn Sina's Critique of Atomism on Subsequent Kalam Discussions of Atomism}, volume={25}, DOI={10.1017/S0957423914000101}, number={1}, journal={Arabic Sciences and Philosophy}, author={Dhanani, Alnoor}, year={2015}, pages={79–104}}

@misc{neumannautomata, title={Theory of Self-Reproducing Automata},author={John Von Neumann},url={https://archive.org/details/theoryofselfrepr00vonn_0/page/n5/mode/2up}, journal={Internet Archive}, year={1966}, language={en} }

@article{causalset,
  title = {Space-time as a causal set},
  author = {Bombelli, Luca and Lee, Joohan and Meyer, David and Sorkin, Rafael D.},
  journal = {Phys. Rev. Lett.},
  volume = {59},
  issue = {5},
  pages = {521--524},
  numpages = {0},
  year = {1987},
  month = {Aug},
  publisher = {American Physical Society},
  doi = {10.1103/PhysRevLett.59.521}
}

@InCollection{Pythagoreanism,
	author       =	{Huffman, Carl},
	title        =	{{Pythagoreanism}},
	booktitle    =	{The {Stanford} Encyclopedia of Philosophy},
	editor       =	{Edward N. Zalta and Uri Nodelman},
	howpublished =	{\url{https://plato.stanford.edu/archives/sum2024/entries/pythagoreanism/}},
	year         =	{2024},
	edition      =	{{S}ummer 2024},
	publisher    =	{Metaphysics Research Lab, Stanford University}
}

@book{platopythag,
    author = {Horky, Phillip Sidney},
    title = "{Plato and Pythagoreanism}",
    publisher = {Oxford University Press},
    year = {2013},
    month = {08},
    isbn = {9780199898220},
    doi = {10.1093/acprof:oso/9780199898220.001.0001}
}

@misc{ObserverTheory—StephenWolframWritings_2023, title={Observer theory}, url={https://writings.stephenwolfram.com/2023/12/observer-theory/}, journal={Stephen Wolfram Writings RSS}, author={Wolfram, Stephen}, year={2023}, month={Dec}}

@article{Tegmark_1998,
   title={The Interpretation of Quantum Mechanics: Many Worlds or Many Words?},
   volume={46},
   ISSN={1521-3978},
   DOI={10.1002/(sici)1521-3978(199811)46:6/8<855::aid-prop855>3.0.co;2-q},
   number={6–8},
   journal={Fortschritte der Physik},
   publisher={Wiley},
   author={Tegmark, Max},
   year={1998},
   month=nov, pages={855–862} }

@book{cartwrightlies,
    author = {Cartwright, Nancy},
    title = "{How the Laws of Physics Lie}",
    publisher = {Oxford University Press},
    year = {1983},
    month = {06},
    isbn = {9780198247043},
    doi = {10.1093/0198247044.001.0001}
}

@article{gorard1, title={Some Quantum Mechanical Properties of the Wolfram Model}, volume={29}, DOI={https://doi.org/10.25088/ComplexSystems.29.2.537}, number={2}, journal={Complex Systems}, author={Gorard, Jonathan}, year={2020}, pages={537–598}}

@article{gorard2,
  author       = {Jonathan Gorard},
  title        = {Some Relativistic and Gravitational Properties of the Wolfram Model},
  journal      = {CoRR},
  volume       = {abs/2004.14810},
  year         = {2020},
  url          = {https://arxiv.org/abs/2004.14810},
  eprinttype    = {arXiv},
  eprint       = {2004.14810},
  bibsource    = {dblp computer science bibliography, https://dblp.org}
}

@article{dirac,
author = {Dirac, Paul Adrien Maurice},
title = {A new basis for cosmology},
journal = {Proceedings of the Royal Society of London. Series A. Mathematical and Physical Sciences},
volume = {165},
number = {921},
pages = {199-208},
year = {1938},
doi = {10.1098/rspa.1938.0053}
}

@book{Tegmark_book, place={New York}, title={Our mathematical universe: My quest for the ultimate nature of reality}, publisher={Vintage Books}, author={Tegmark, Max}, year={2015}}

@InCollection{Blanchette_2007,
	author       =	{Blanchette, Patricia},
	title        =	{{The Frege-Hilbert Controversy}},
	booktitle    =	{The {Stanford} Encyclopedia of Philosophy},
	editor       =	{Edward N. Zalta and Uri Nodelman},
	howpublished =	{\url{https://plato.stanford.edu/archives/spr2024/entries/frege-hilbert/}},
	year         =	{2024},
	edition      =	{{S}pring 2024},
	publisher    =	{Metaphysics Research Lab, Stanford University}
}

@misc{Chojna_Warendorff_2023, title={Wolfram Language paclet for computational analysis of chess games}, url={https://resources.wolframcloud.com/PacletRepository/resources/Wolfram/Chess/}, journal={Wolfram/Chess}, author={Chojna, Rafal and Warendorff, Jay}, year={2023}}

@misc{metaphysicsaristotle,title={The Internet Classics Archive | Metaphysics by Aristotle}, url={http://classics.mit.edu/Aristotle/metaphysics.14.xiv.html}}

@book{Heisenberg, title={Physics and philosophy}, author={Heisenberg, Werner}}

@InCollection{aristotle,
	author       =	{Cohen, S. Marc and Reeve, C. D. C.},
	title        =	{{Aristotle’s Metaphysics}},
	booktitle    =	{The {Stanford} Encyclopedia of Philosophy},
	editor       =	{Edward N. Zalta},
	howpublished =	{\url{https://plato.stanford.edu/archives/win2021/entries/aristotle-metaphysics/}},
	year         =	{2021},
	edition      =	{{W}inter 2021},
	publisher    =	{Metaphysics Research Lab, Stanford University}
}

@misc{Tegmark, title={Critique}, url={https://space.mit.edu/home/tegmark/mathematical.html}, journal={The universes of max tegmark}, author={Tegmark, Max}}

\end{document}